# Magnetic nanostructures by adaptive twinning in strained epitaxial films

Sandra Kauffmann-Weiss[1,2], Markus E. Gruner[3,*], Anja Backen[1,2], Ludwig Schultz[1,2], Peter Entel[3] and Sebastian Fähler[1]

[1] IFW Dresden, PO Box 270116, 01171 Dresden, Germany

[2] Dresden University of Technology, Institute for Materials Science, 01062 Dresden, Germany

[3] Faculty of Physics and Centre for Nanointegration, CeNIDE, University of Duisburg-Essen, D-47048 Duisburg, Germany

**We exploit the intrinsic structural instability of the $Fe_{70}Pd_{30}$ magnetic shape memory alloy to obtain functional epitaxial films exhibiting a self-organized nanostructure. We demonstrate that coherent epitaxial straining by 54% is possible. The combination of thin film experiments and large-scale first-principles calculations enables us to establish a lattice relaxation mechanism, which is not expected for stable materials. We identify a low twin boundary energy compared to a high elastic energy as key prerequisite for the adaptive nanotwinning. Our approach is versatile as it allows to control both, nanostructure and intrinsic properties for ferromagnetic, ferroelastic and ferroelectric materials.**

Straining the crystal lattice by coherent epitaxial growth is a well known approach adjusting functional properties in semiconductor, ferromagnetic, ferroelectric and multiferroic films. However, the high elastic energy due to the misfit between substrate and film limits maximum strain and critical thickness which can be achieved with this approach. These limitations can be overcome when using materials with a structural instability [1]. The disordered ferromagnetic shape memory alloy $Fe_{70}Pd_{30}$ [2], for example, shows a martensitic transition that follows the Bain transformation path [3] from face centered cubic (*fcc*) to body centered cubic (*bcc*) structure. Since the energy for straining $Fe_{70}Pd_{30}$ is relatively low, different tetragonal distortions along the Bain path could be stabilized. This allows tuning intrinsic magnetic properties like Curie temperature, orbital magnetic moment and magnetocrystalline anisotropy [1]. For functional magnetic materials, however, excellent extrinsic properties are required in addition. For instance, percolated media as a candidate for the next generation of magnetic recording media require a substantial increase of pinning centers [4,5]. These depend on an appropriately designed defected microstructure, which is not present in coherently strained films.

---

[*] E-mail: Markus.Gruner@uni-due.de

In this letter, we explain how appropriate nanostructures can be obtained by pushing the limits of coherent epitaxial growth beyond the stability range of the known bulk phases. We exploit a novel relaxation mechanism resulting in the formation of self-organized nanostructures, which is closely related to structural ferroelastic instabilities, while keeping the freedom to tune the intrinsic magnetic properties. We identify volume elastic energy and twin boundary interface energy as competing contributions. This concept is universally applicable for various functional materials, including ferroelectrics.

As a prototype material exhibiting structural instabilities [6], we selected the chemically disordered $Fe_{70}Pd_{30}$ alloy. In vicinity of the alloy composition $Fe_{70}Pd_{30}$ various tetragonal phases are observed in bulk, which are connected by diffusionless martensitic transformations [7]. According to the concept of a Bain transformation, these phases can be identified as intermediate stages between the *fcc* austenite ($c/a = 1$) and *bcc* martensite ($c/a = \sqrt{1/2}$). The tetragonal distortion $c/a$ with respect to the *fcc* unit cell is the key parameter for the description of the various phases. In our previous work [1], we grew a series of $Fe_{70}Pd_{30}$ films on different substrate materials using sputter deposition at room temperature. The substrate defines different in-plane lattice constants, which allowed varying the tetragonal distortion within the Bain transformation path. In the thickness range from 50 nm up to 2 μm no relaxation was observed [8]. In the present work, we introduce Cu as a new substrate, which induces a tetragonal distortion of $c/a = 1.09$ within the $Fe_{70}Pd_{30}$ film – a value well beyond the Bain transformation path.

To probe epitaxial growth and its limits we used (111) pole figure measurements of films with different thicknesses (Fig. 1, for experimental details see the supplemental material). For a 40 nm thick film the pole figure (Fig. 1a) is dominated by 4 reflections at a tilt angle $\Psi$ of 57° (the broadening around the rotation angle $\Phi$ is discussed later). Together with the pole figure measurements of the Cu substrate (not shown) this reveals an epitaxial relationship of FePd(001)[100]||Cu(100)[001]. The connection of the angle $\Psi$ with the $c/a$ ratio in a tetragonal crystal [1] gives a $c/a$ ratio of 1.09. As an independent probe, we measured the out-of-plane lattice constant $c = 0.397$ nm in Bragg Brentano geometry and the (111) lattice spacings in tilted geometry, which give an in-plane lattice constant of $a = 0.362$ nm, close to $a_{Cu} = 0.3615$ nm. Thus the in-plane lattice parameter of the $Fe_{70}Pd_{30}$ layer is fixed by the interface to the Cu substrate. This experiment shows that we can coherently strain $Fe_{70}Pd_{30}$ films beyond the Bain path and vary the tetragonal distortions in a wide range of 54% with respect to the bcc ground state. Even beyond the Bain path all data points can be described by a constant unit cell volume (Fig. 2a).

In order to understand the influence of the elastic energy associated with coherent epitaxial growth from first principles, we simulate a 500 atom $Fe_{68}Pd_{32}$ supercell using density functional theory (DFT) description of the electronic structure [9]; for details see Ref. 10. To realize an equivalent constraint as in epitaxial films we fix both in-plane axes of the box $a_{box}$, while the out-of-plane axis $c_{box}$ is determined by the constant unit cell volume. Previous DFT work considered chemical disorder within an effective medium approach (coherent potential approximation, CPA) [1,11], which requires to fix the atoms to the ideal tetragonal lattice positions. One obtains an energy landscape $E(c/a)$ between the *fcc* and *bcc* lattice which is essentially flat compared to thermal energies (blue squares in Fig. 2b, 1 meV≙11.6 K). This suggests that coherent epitaxial growth of thick films should be possible *within* the Bain path. Outside the Bain path total energy increases indicating that the critical thickness of the films should be low. However, allowing for deviations from the ideal lattice positions considerably reduces the total energy (red circles in Fig. 2b) and changes the energy landscape of the disordered system substantially [10]. These deviations originate from accommodations of the interatomic forces, which arise, e.g., from the different size of the atoms. In agreement with bulk experiments [7] now a *bcc* ground state is observed whereas for the ideal lattice the energy is minimized at *fcc*. This qualitative change is attributed to the reduced coordination in the open *bcc* structure which imposes fewer constraints for the relaxation. The relaxations lead furthermore to a second local minimum beyond the Bain Path, at about $c/a$ = 1.08. This is close to the experimentally realized value of 1.09 and suggests that our film is in a metastable state.

The energy difference with respect to the ground state is the driving force for the lattice relaxation, which commonly proceeds by the introduction of misfit dislocations or deformation twins [12]. A critical thickness limiting coherent epitaxial growth occurs as the total elastic energy increases proportionally with film thickness. The defect energy associated with the relaxation process, however, is required only once and allows continuing further film growth in the relaxed ground state. Hence the critical thickness is determined by the balance of both energies.

Relaxation is evident from pole figure measurements of films with increased thickness (Fig. 1). The substantial differences observed for films with 40, 300 and 2000 nm thickness imply a two stage relaxation mechanism. In the thickness range between 40 and 300 nm a transition from four rather sharp peaks (Fig. 1a) to significantly broadened ones (in particular around Ψ) is observed. This behavior is neither expected for misfit dislocations nor common deformation twinning, but suggests a different mechanism involving a rotation around the substrate normal.

Our DFT calculations allow to monitor the first relaxation process with atomic resolution (inset in Fig. 2b, a video is available online). It proceeds spontaneously and reversibly on the shallow energy landscape. Relaxation results in two equally sized twins within the simulation box, which are connected by atomically sharp twin boundaries with $(101)_{fct}$ orientation. The projection shown in Fig. 2b is identical to a top view of the film and suggests that the experimentally observed broadening around Ψ originates from the bending at a twin boundary. This spontaneous pattern formation has striking similarities with the formation of nanotwins within the adaptive martensite [13] of the prototype magnetic shape memory alloy $Ni_2MnGa$ [14]. A modified geometrical concept of adaptive martensite allows identifying the relevant energy contributions for this spontaneously formed nanostructure. In this concept elastic energy is reduced on cost of increased twin boundary energy. The total energy curve (open circles in Fig. 2b) suggests that reduction of elastic energy is possible when the tetragonality of the nanotwins, $c/a|_{\text{twin}}$, approaches √1/2. In order to keep compatibility with the simulation box, the twins orient in a way that one long $a|_{twin}$ axis of the nanotwins is parallel and commensurate to $c|_{box}$. Along $a|_{box}$ an alternating arrangement of $a|_{twin}$ and $c|_{twin}$ form twins. Elementary geometry connects both tetragonal distortions by

$$\left.\frac{c}{a}\right|_{box} = \sqrt{\frac{1}{2} + \frac{1}{2}\left(\left.\frac{c}{a}\right|_{twin}\right)^{-2}}. \tag{1}$$

A small mismatch of the atoms at the edges of adjacent cells in second order in $1-c/a|_{twin}$ is superseded by the disorder-related atomic relaxations (see supplemental material). There is no degree of freedom; hence the selection of the substrate lattice constant uniquely determines the lattice constants of the twin. To illustrate that this process can indeed reduce elastic energy, we plotted the associated elastic energy $E(c/a|_{twin})$ as dashed line in Fig. 2b. According to [13, 14] the energy difference to the total energy $E(c/a|_{box})$ allows calculating the twin boundary energy $\gamma_{twin}$:

$$\gamma_{twin} = \left(E\left(\left.\frac{c}{a}\right|_{box}\right) - E\left(\left.\frac{c}{a}\right|_{twin}\right)\right)\bigg/ A_{TB} \tag{2}$$

$A_{TB}$ is the area of both $(101)_{fct}$ twin boundaries within the simulation box. Considering that shear energy increases with deviation from the *fcc* austenite ($c/a|_{twin} = 1$), the observed dependency of $\gamma_{twin}$ (Fig. 3a) agrees well with the quadratic increase expected from linear elastic theory. The absolute values of $\gamma_{twin}$ are comparably small as those reported for the Ni-Mn-Ga [14] and NiTi [15] shape memory systems. Furthermore, the geometric analysis of Khachaturian et al. [13] shows that adaptive phases can exist in the $Fe_{70}Pd_{30}$ system, which requires low twin boundary energy as an essential prerequisite.

Their formation reduces total energy since the reduction in elastic energy exceeds the defect energy required for the formation of the twin boundaries. The periodicity, however, is imposed by the finite size of the simulation box and may be larger for films. However, such a coarsening process increases the misfit to the underlying substrate lattice and consequently increases elastic energy. If we consider the defect energy associated with the coarsening process, the nanotwinned configuration is often retained as a metastable phase [16]. Though being unable to predict the exact periodicity, we consider the selected simulation boundaries as an appropriate first approximation to describe the conditions of a film constrained by an epitaxial interface to the substrate. In particular it gives the appropriate orientation of the twin boundary perpendicular to the substrate, as all other orientations are inhibited since they would involve a bending of the rigid substrate [16].

Despite the striking similarities to bulk adaptive martensite, thin film adaptivity exhibits important differences. In bulk, it is formed due to the elastic constraints at the habit plane between austenite and martensite [13], which has irrational indices. In the present thin film experiment the interface to the substrate realizes an artificial (001) constraint. While in bulk the martensitic transformation proceeds within the Bain path, coherent epitaxial growth allows to go beyond. Therefore, substantially different behavior can be observed for $c/a < 1$ and $c/a > 1$. When taking austenite ($c/a = 1$) as a starting point, the growth on a substrate with $c/a < 1$ already results in a reduction of energy (open circles in Fig. 2b). There is no need to introduce twin boundaries; equivalent twinning implies $c/a|_{twin} > 1$ and would thus increase elastic energy. On a substrate with $c/a > 1$, adaptive nanotwinning reduces the total energy. The observed local energy minimum allows considering this microstructure as metastable adaptive phase. This asymmetry explains why only biaxial *compressive* stress ($c/a > 1$) induces an adaptive phase as a relaxation mechanism.

The relaxation mechanism involves $(101)_{fct}$ transformation twinning and is thus different to the common introduction of misfit dislocations or (111) deformation twinning of *fcc* films [12]. In a cubic system a (101) twin boundary transforms the crystal into its identity, thus the $(101)_{fct}$ twinning system is only present in crystals having a reduced symmetry. Hence the structural instability is an essential precondition for the relaxation mechanism. This can be linked to the presence of a TA phonon instability which is frequently observed in the vicinity of martensitic transformations [17], also in $Fe_{70}Pd_{30}$ [18]. The phonon softening describes the same (infinitesimal) shear deformation occurring at a macroscopic twin boundary, providing an experimental indication for very low interface energy of the transformation twins.

To confirm the continuum interpretation of adaptive twinning we extract the pair distribution function $g(r)$ from the relaxed atomic configuration at each $c/a|_{box}$ (Fig. 4), representing the probability to find two atoms at a given distance *r*. It provides a unique fingerprint which allows discriminating differently coordinated structures. For comparison neighbor distances calculated for an ideal tetragonal distortion are plotted as lines. Without relaxation ($c/a|_{twin} = c/a|_{box}$, solid lines) a good match is obtained within the Bain transformation path ($c/a|_{box} < 1$), but this description fails beyond the Bain path. Instead, the neighbor distances expected from a relaxed tetragonal lattice according to Eq. (1), yield an appropriate agreement (dotted line). Indeed, the tetragonal distortion $c/a|_{twin}$ obtained by a least square fit to $g(r)$ (Fig. 3b) agrees well with the geometric prediction given by Eq. (1). This confirms that only minor deviations of the lattice configuration from a simple, sharp twin boundary configuration occur.

The fixed geometric relations between $c/a|_{box}$ and $c/a|_{twin}$ plotted in Fig. 3b, do not allow a transformation towards the *bcc* ground state which could further reduce elastic energy. Hence, a further lattice relaxation mechanism is required to approach the ground state. The pole figure measurements of 2000 nm thick films reveal a (211) fiber texture, which suggests that (211) deformation twinning occurs as second relaxation mechanism, which is the favored deformation twinning system for a *bcc* lattice.

In conclusion, the present combination of thin film experiments and density functional theory allows identifying adaptive nanotwinning as a novel relaxation mechanism in coherently strained epitaxial films. Our analysis implies that it is appropriate to use the continuum concepts of elastic energy and twin boundary energy to describe the spontaneous formation of these nanostructures. For materials exhibiting structural instabilities, these energies are closely related to the transformation path and phonon instability, respectively. In magnetic shape memory alloys, both depend on temperature and composition [19,20] and we expect that this will allow to control the feature size. Selection of the substrate lattice spacings uniquely determines the tetragonal distortion of the relaxed lattice and hence supplies the freedom to tailor both, its microstructure and anisotropic magnetic properties. This approach is applicable to explain pattern formation in other functional materials exhibiting structural instabilities. Similar to magnetic shape memory alloys [14], low symmetric adaptive phases exhibiting the best functional properties exist in ferroelectric materials like PMN-PT [21]. As these metastable phases can equally be identified as nanotwinned microstructures the present concepts can explain the observed asymmetry between tensile and compressive stress in thin ferroelectric films [22]. We expect that our identification of elastic and twin boundary energy as key competing energies will also pave the way towards better understanding of nano chessboards [23], which may involve spinodal decomposition [24].

The authors thank M. Richter, A. Diestel and R. Niemann for discussions and experimental support and the DFG for funding via the Priority Program SPP 1239. They also thank the John von Neumann Institute for Computing, the Jülich Supercomputing Center (NIC and JSC, both Forschungszentrum Jülich) and the Center for Computational Sciences and Simulation (CCSS, University of Duisburg-Essen) for computing time and support.

Figures:

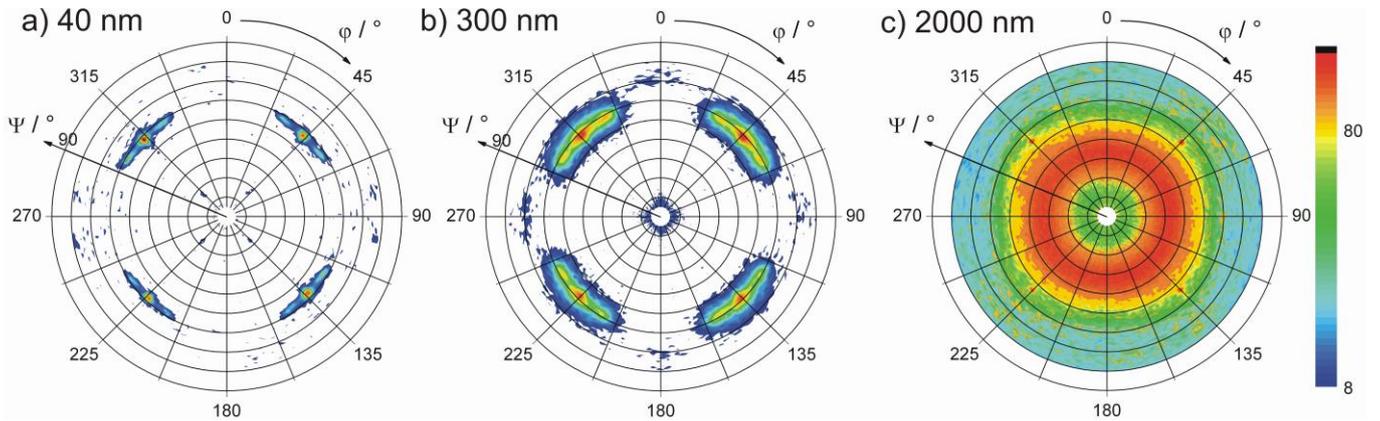

**Fig. 1:** Probing coherent epitaxial growth and relaxation mechanisms of $Fe_{70}Pd_{30}$ films on a Cu (100) substrate. (a) The fourfold symmetry of the (111) pole figure at $\Psi = 57°$ of the 40 nm thick film verifies the epitaxial growth. (b) For the 300 nm thick film a broadening in $\Psi$ and $\Phi$ direction of the four (111) peaks is observed. (c) In 2000 nm thick films these peaks disappear and a (211) fiber texture is visible.

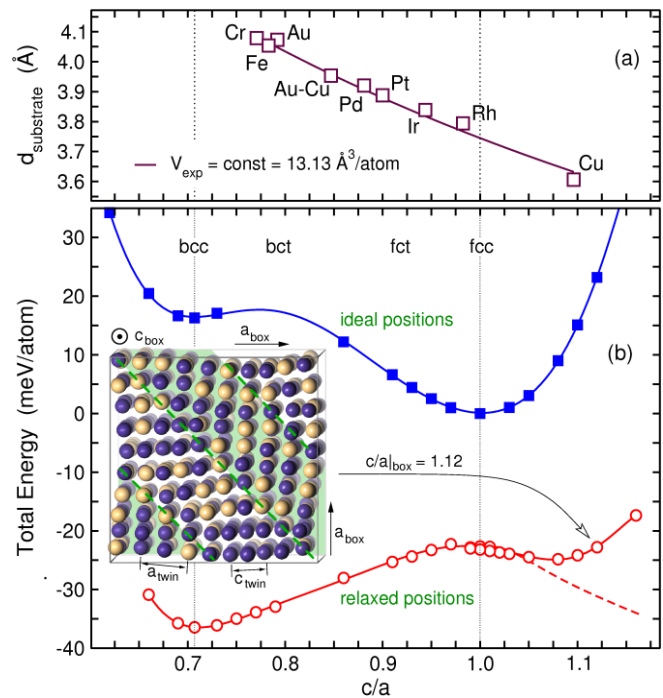

**Fig. 2:** (a) Film tetragonality $c/a$ obtained after growth on various substrate materials, exhibiting different lattice spacing $d_{substrate}$. The volume per atom remains nearly constant in the films. (b) Total Energy at different tetragonal distortion $c/a$ from *ab initio* calculations. When fixing the atoms at their ideal lattice positions, the blue curve with an *fcc* global energy minimum is obtained. Allowing deviations from these positions results in a substantial reduction in total energy (red curve) and a *bcc* ground state. A further local minimum is found at $c/a = 1.08$. As described in the text, the red dashed curve describes the contribution from elastic energy only according to Eqs. (1) and (2). The inset exemplarily depicts the atomic configuration of a relaxed configuration (Fe: blue, Pd: yellow) in a simulation box with $c/a|_{box} = 1.12$. The perspective is equivalent to a top view

on the film. Adaptive nanotwinning occurs as a collective relaxation mechanism, the twin boundaries are marked as dashed green lines.

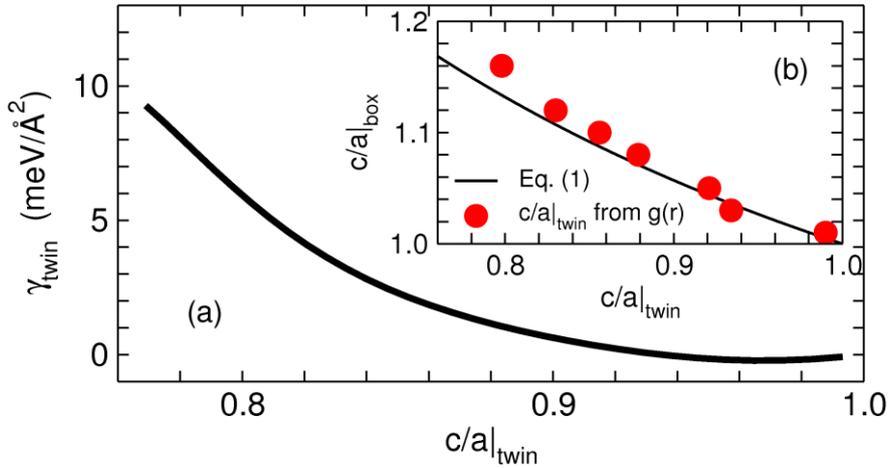

**Fig. 3:** (a) The twin boundary energy $\gamma_{twin}$ for the relaxed lattice using Eq. (2) increases with tetragonal distortion of the twin $c/a|_{twin}$. (b) Correlation of the tetragonal distortion of the *ab initio* box $c/a|_{box}$ representing the epitaxial constraint and the resulting tetragonal distortion of the twin $c/a|_{twin}$. The solid line represents the geometrical constraint, Eq. (1), while the red points are derived from a least mean square fit of the pair distribution function of the twinned structures to the best matching untwinned configuration with $c/a|_{box}<1$.

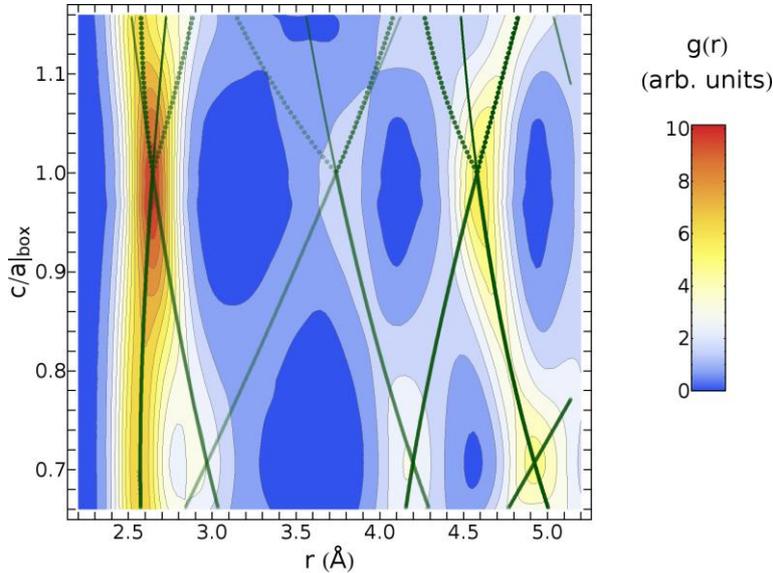

Fig. 4: Contour plot of the radial pair distribution function $g(r, c/a|_{box})$ for different tetragonal distortions of the simulation box $c/a|_{box}$. For positive and negative deviations from $c/a|_{box}=1$ similar atomic coordination are observed. The solid lines refer to the neighbor distances of an ideal tetragonal structure without lattice relaxations ($c/a|_{twin}=c/a|_{box}$). While this is an appropriate description for $c/a|_{box}<1$, it fails beyond the Bain path. In this region a relaxed nanotwinned structure with $c/a|_{twin}$ according to Eq. (1) describes the atomic configuration well (dotted lines). The opacity of the lines corresponds to the number of neighbor pairs in the respective coordination shell.